\documentclass[twopage,11pt] {article}

\usepackage[dvips]{graphicx}

\begin{document}

\title{Cosmic Ray Astrophysics with AMS-02}

\author{Diego Casadei\\ \small
\emph{INFN, Sezione di Bologna}\\ \small
Via Irnerio 46, 40126 Bologna, Italy}

\date{April 27, 2004}

\maketitle

\pagestyle{myheadings} 
\thispagestyle{plain}         
\markboth{Diego Casadei}{Cosmic Ray Astrophysics with AMS-02}
\setcounter{page}{1}         


 The Alpha Magnetic Spectrometer (AMS) is a cosmic ray (CR) experiment
 that will operate on the International Space Station for three years,
 measuring the particle spectra in the rigidity range from 0.2 GV to 2
 TV.  The AMS-02 detector will provide measurements with unprecedented
 statistics of the hadronic and leptonic cosmic rays, allowing for a
 better study of the Earth magnetosphere through the secondaries
 produced by CR interactions in the atmosphere; of the solar system
 environment through the measurement of the solar modulation over a
 long period; of the solar system neighborhood through the measurement
 of the ratio between unstable isotopes and stable elements; of the
 interstellar medium of our Galaxy through the ratio between secondary
 and primary isotopes and the measurement of proton and helium
 spectra.


\section{Introduction}  

 The Alpha Magnetic Spectrometer (AMS) is a cosmic ray (CR) experiment
 that will operate on the International Space Station (ISS) for three
 years, measuring the particle spectra in the rigidity range from 0.2
 GV to 2 TV.  The first version of the detector, called AMS-01, was
 flown on board of the space shuttle Discovery in 1998, from June 2 to
 12.  The second version of the detector, called AMS-02, will operate
 on the ISS starting from year 2007.  AMS-01 collected more than
 hundred million events, and the results were published about the CR
 protons and antiprotons, helium, electrons and positrons, and the
 upper limit on the antimatter ratio (for a review, see
 Ref.~\cite{ams01}).

\begin{figure}[t]
\includegraphics[width=0.44\textwidth]{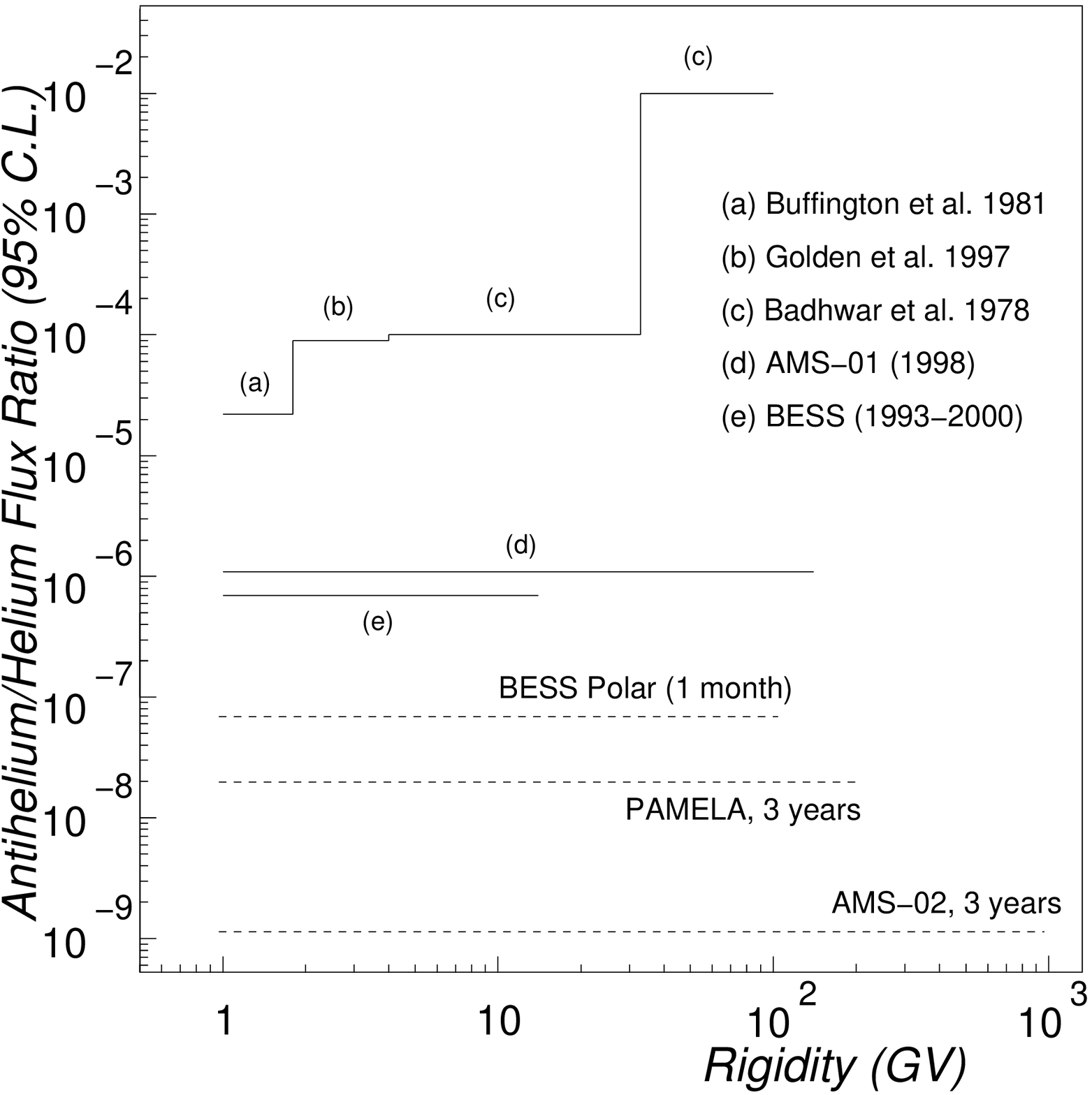} \hfill
\includegraphics[width=0.48\textwidth]{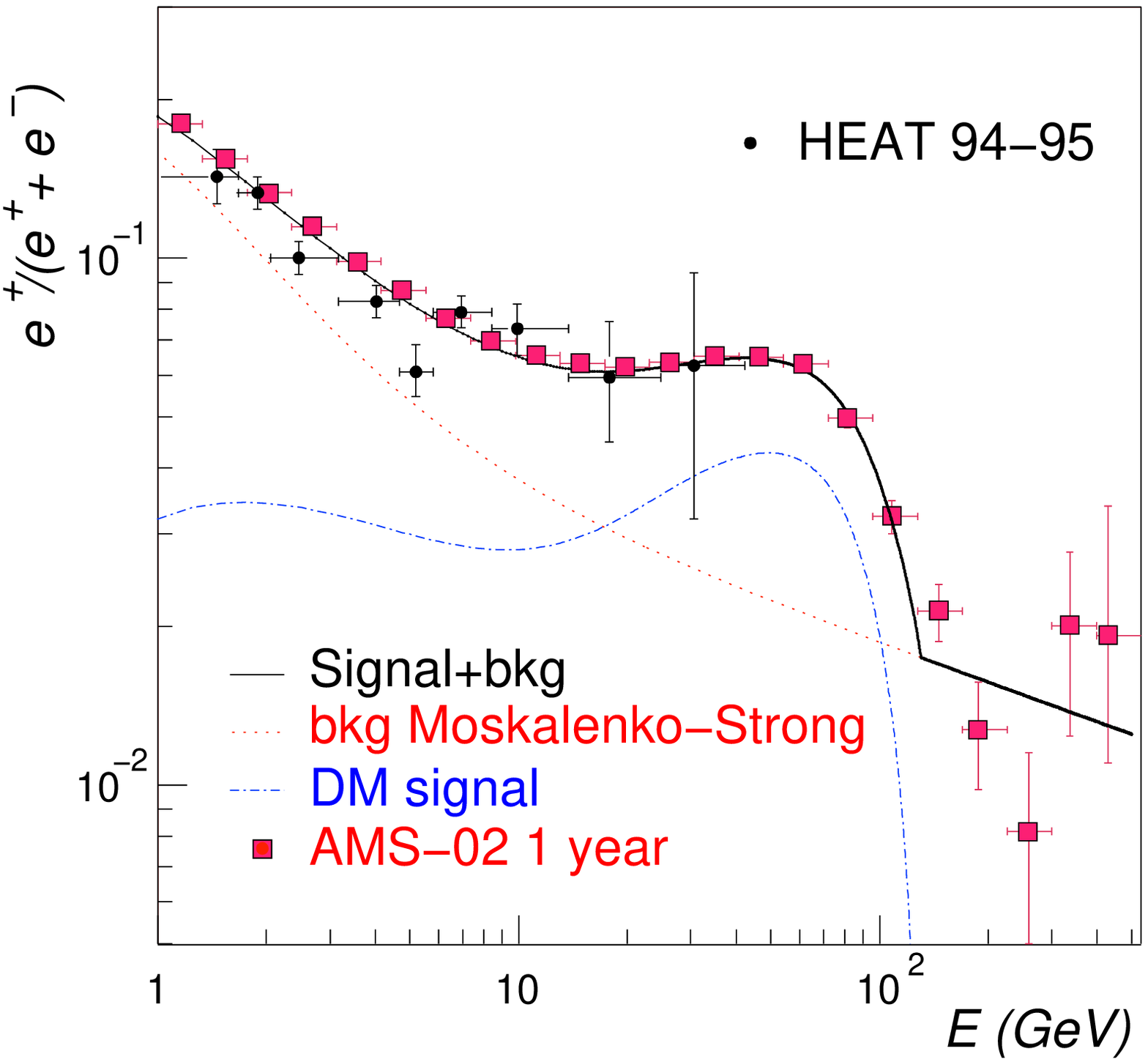}
\caption{Upper limits on the CR antihelium to helium ratio (left
  panel), and a possible signature of the neutralino annihilation in
  the ratio between CR positrons and electrons (right
  panel).}\label{fig1}
\end{figure}

 The physics goals of the AMS experiment are: (1) the search for
 antinuclei among the CR, that would be a smoking gun in favor of the
 existence of cosmic antimatter in form of anti-stars
 (figure~\ref{fig1}, left panel, shows the upper limits on the
 antihelium to helium ratio);
 (2) the search for signatures of the supersymmetric dark matter
 candidates, looking at the antiproton and positron spectra
 (figure~\ref{fig1}, right panel, shows the expected distortion of the
 positron to electron ratio that would be produced by the annihilation
 of 65 GeV/$c^2$ neutralinos \cite{lamanna03});
 (3) the high-statistics study of the CR
 particle spectra up to 1--2 TV rigidity.

\begin{figure}[t!]
\centering
\includegraphics[width=0.4\textwidth]{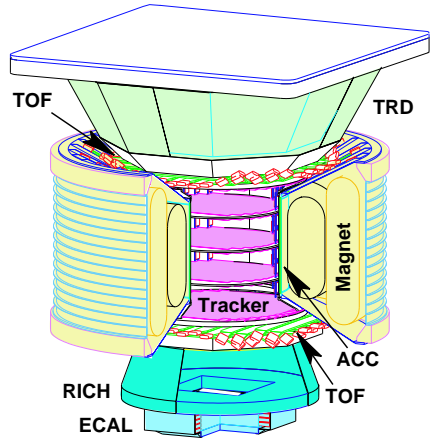}
\caption{The AMS-02 detector.}\label{fig2}
\end{figure}

 The AMS-02 detector (figure~\ref{fig2}) is able to make multiple
 measurements of the energy lost by the particles traversing it, of
 their track and of their velocity \cite{ams02}.  The basic technique
 for particle recognition lays on the measurement of the energy loss,
 giving the particle charge absolute value, of the track curvature,
 giving the particle rigidity, and of the velocity value and
 direction, that gives the charge sign together with the curvature
 measurement.

 The anticoincidence system (ACC) vetoes tracks crossing the magnet,
 hence all particles must traverse the transition radiation detector
 (TRD), the time of flight system (TOF), the silicon tracker and the
 proximity focusing \v{C}erenkov detector (RICH).  About half of them
 will also produce a shower in the electromagnetic calorimeter (ECAL),
 placed below the RICH.  RICH and TOF are used to distinguish
 electrons and positrons from antiprotons and protons down to 0.2 GeV,
 whereas TRD and ECAL allow for the e/p separation up to 300 GeV, that
 is the upper energy limit for the antiproton and positron
 measurements.  Nuclei can be separated from antinuclei up to the
 maximum detectable rigidity, i.e.~up to 2 TV.  Finally, gamma rays
 can be detected in the largely unexplored energy range above few tens
 of GeV.

 This paper focuses on the ``ordinary'' CR astrophysics items that can
 be studied with the help of the AMS-02 detector.  They are necessary
 steps before anyone will trust the possible exotic results of the
 measurement, but they are also important for the understanding of our
 Galactic environment.

\section{Cosmic Ray Astrophysics}


\begin{figure}[t]
\includegraphics[width=\textwidth]{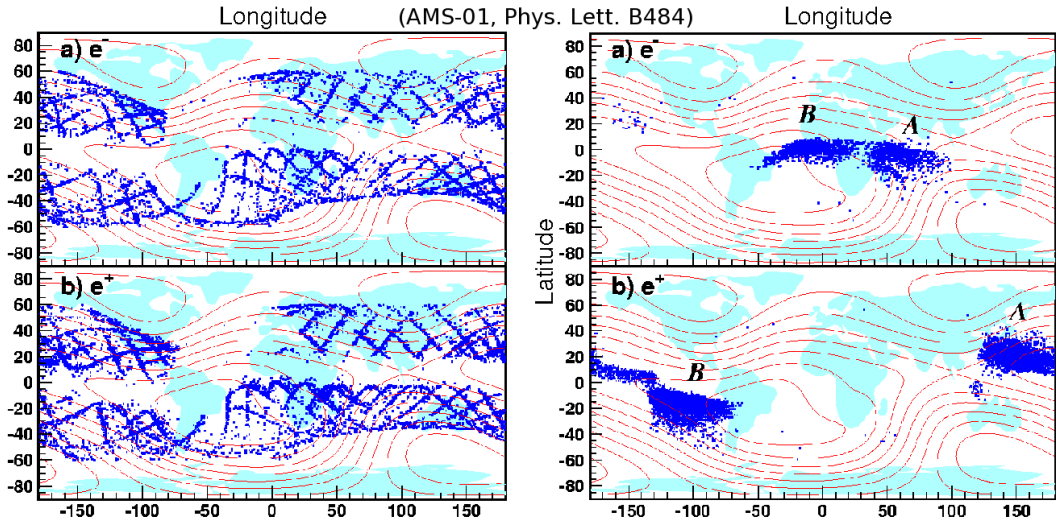}
\caption{The trapped electrons and positrons measured by AMS-01.}\label{fig3}
\end{figure}

\subsection{Earth Magnetosphere}

 Figure \ref{fig3} shows the trapped electrons and positrons measured
 by AMS-01, as function of the geographic coordinates.  The orbit
 inclination was 51.7$^\circ$ and the altitude ranged between 350 and
 390 km.  Field countours at 380 km are also shown.  Back-tracing
 every event, all particles started from the Earth atmosphere were
 considered secondary \cite{ams01el}.  Secondary electrons and
 positrons were detected after a time $\tau$ since their creation, and
 it was found that they can be divided into two populations:
 ``short-lived'' e+/e$-$ ($\tau<0.1$ s, left panel) originated
 uniformly (one can follow the AMS-01 orbits in the left panel),
 whereas ``long-lived'' e+/e$-$ ($\tau>1$ s, right panel) were
 bouncing between two mirror points.  This is due to the kinematics of
 charged particles in the asymmetric geomagnetic field, that was
 sampled by a detector moving between geomagnetic latitudes from
 roughly 75$^\circ$ North to 75$^\circ$ South.  Hence, an orbiting
 detector is a powerful instrument to study the trapped particles in
 the geomagnetic field.

\begin{figure}[t]
\includegraphics[width=0.45\textwidth]{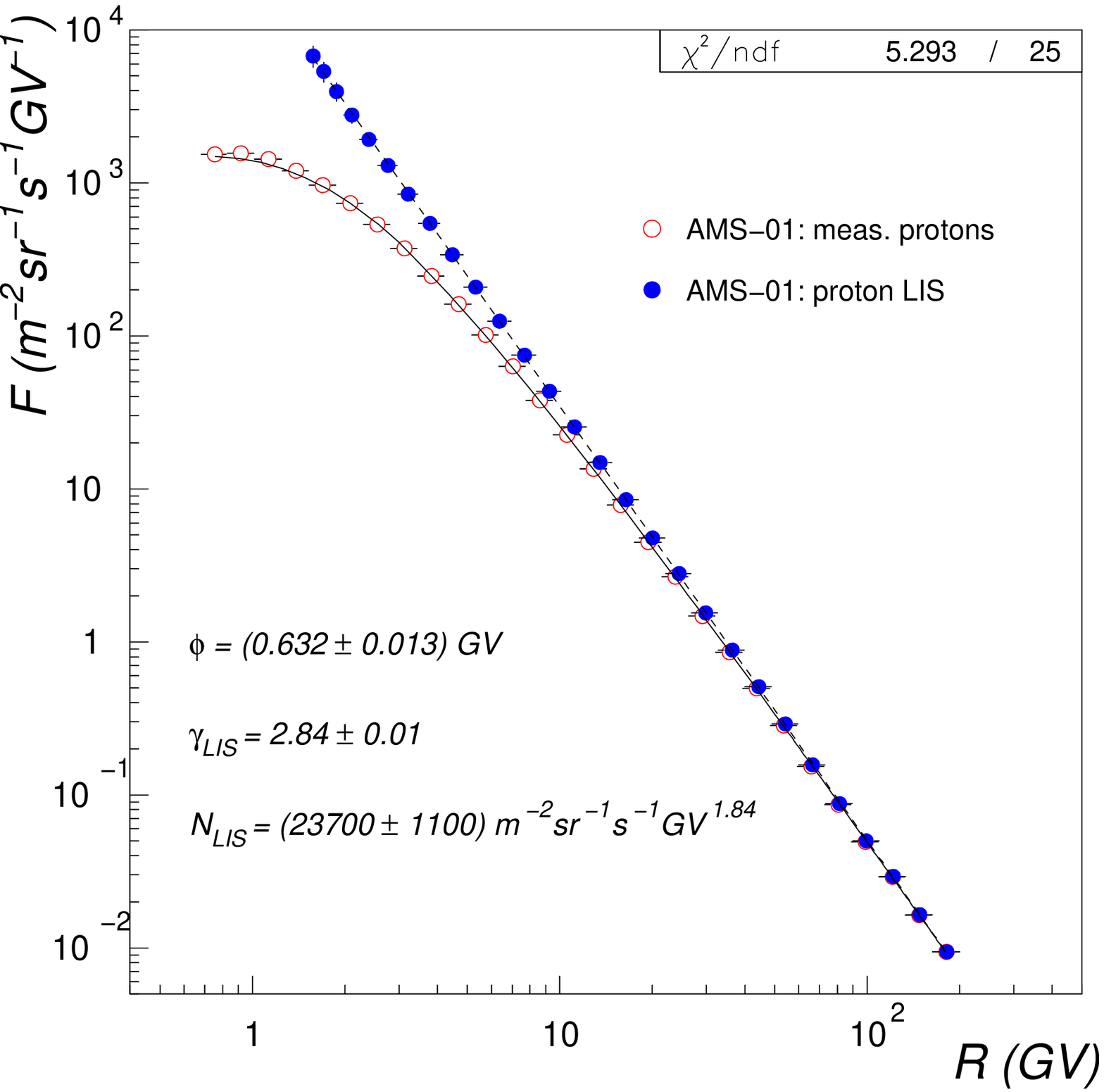} \hfill
\includegraphics[width=0.48\textwidth]{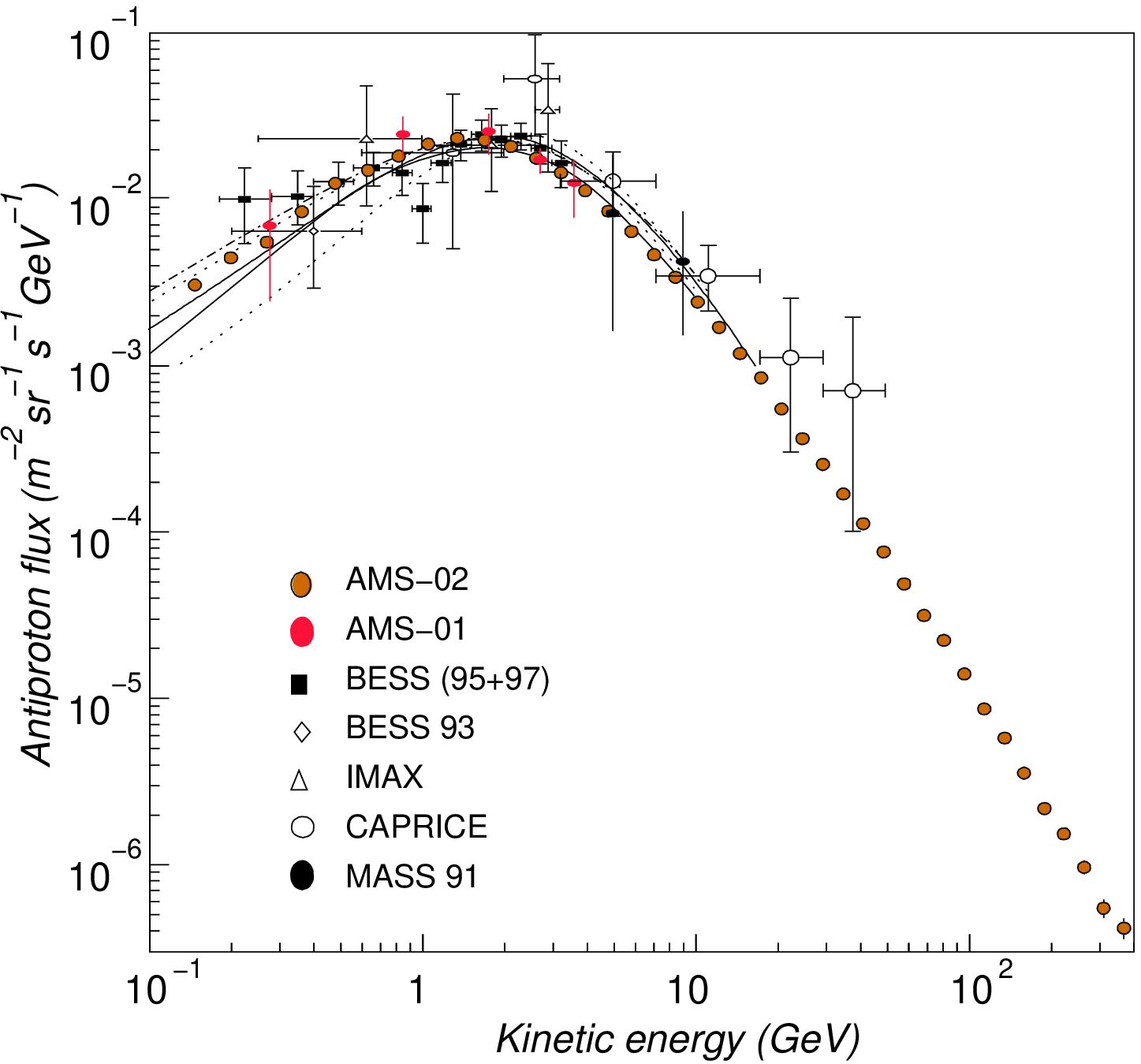}
\caption{The protons measured by AMS-01 can be used to measure the
  solar modulation parameter (left panel).  The antiproton to proton
  ratio is also sensitive to the solar modulation (right
  panel).}\label{fig4}
\end{figure}

\subsection{Solar Modulation}

 The solar activity, parametrized with the number of sun spots or with
 the value of the interplanetary magnetic field, is anticorrelated
 with the total CR flux on the top of the atmosphere, as measured for
 example with high mountain neutron detectors.  The Sun affects the
 diffusion of charged particles below several GeV/nucl following its
 22-years cycles (\emph{solar modulation}).  Figure \ref{fig4}, left
 panel, shows the solar modulation first-order effect on the proton
 spectrum measured by AMS-01 \cite{casadei03}, that it is expected to
 be a single power-law in rigidity $R = p c / (Ze)$ down to hundred
 MV.  At the next order, the solar modulation has a charge-sign
 dependent effect that can be best seen in the antiparticle/particle
 ratio, as shown in the right panel with antiprotons.  This plot shows
 also the expected statistics for AMS-02, that has the potential to
 discover high energy bumps that could be produced by exotic sources,
 like the annihilation of neutralinos, the supersymmetric candidates
 for the dark matter \cite{lamanna03}.

\begin{figure}[t]
\includegraphics[width=0.45\textwidth]{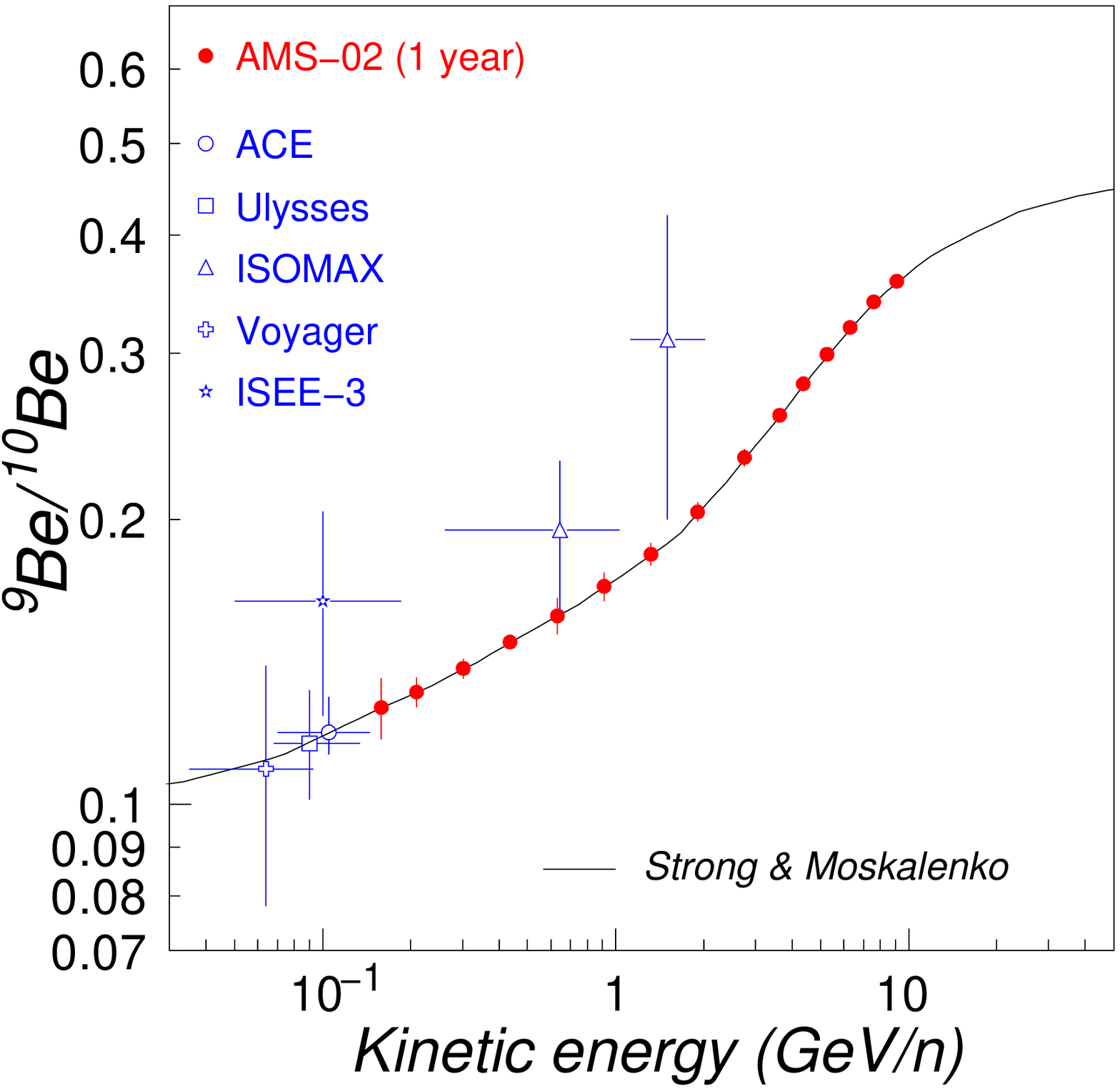} \hfill
\includegraphics[width=0.45\textwidth]{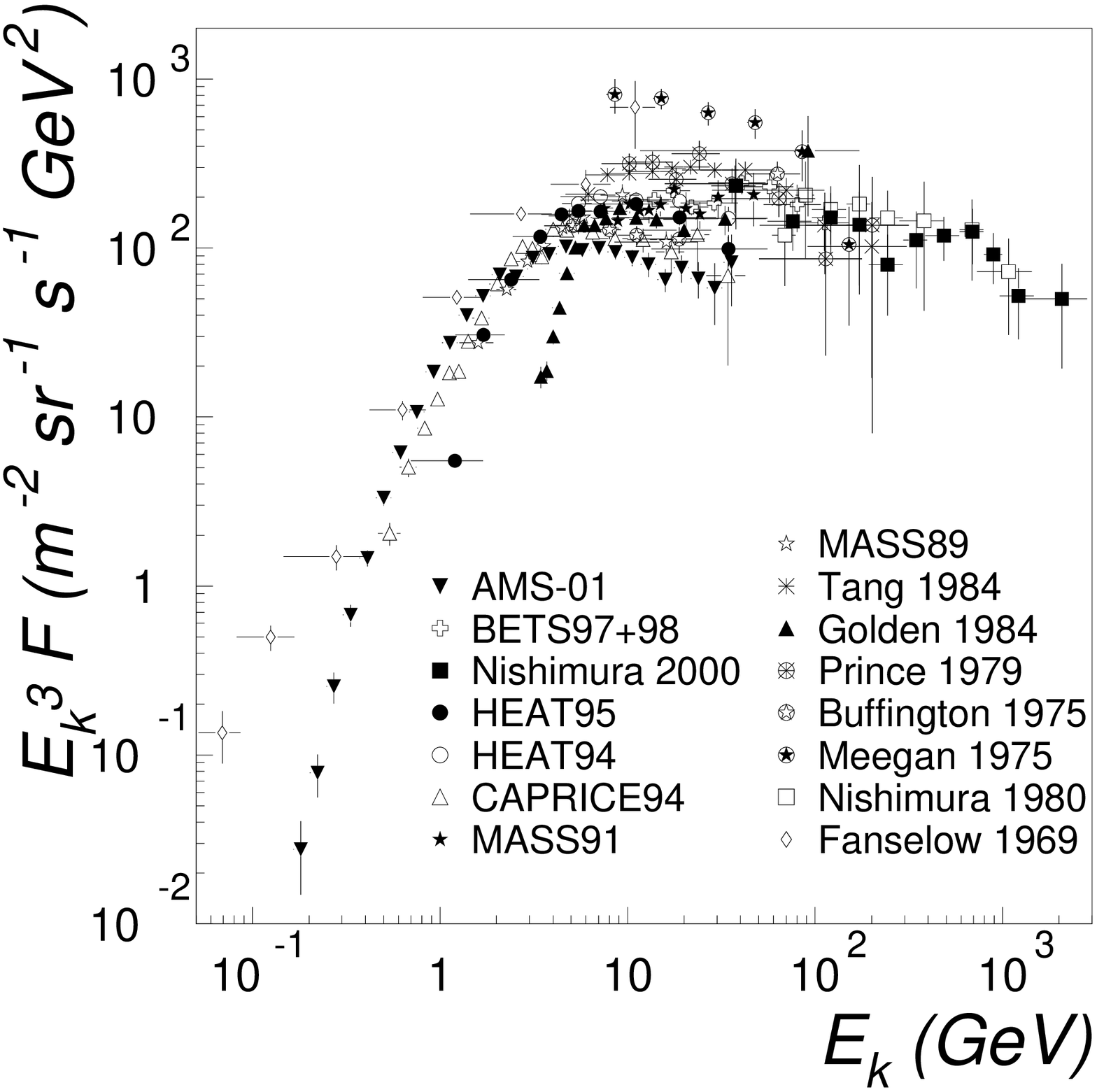}
\caption{Beryllium (left) and electron (right) measurements.  The
  expected AMS-02 statistics is shown in the left panel.}\label{fig5}
\end{figure}

\begin{figure}[t!]
\includegraphics[width=0.46\textwidth]{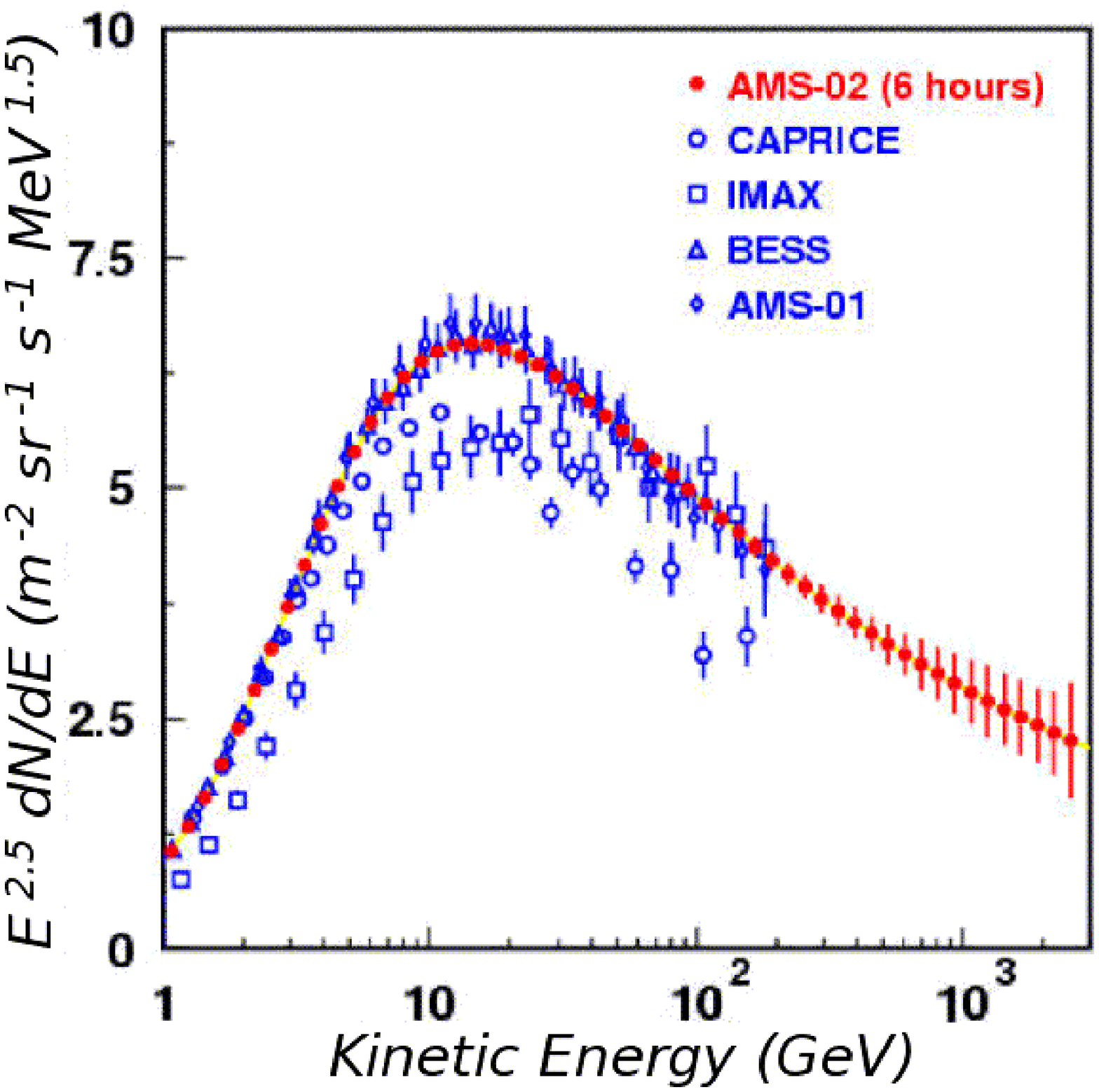} \hfill
\includegraphics[width=0.455\textwidth]{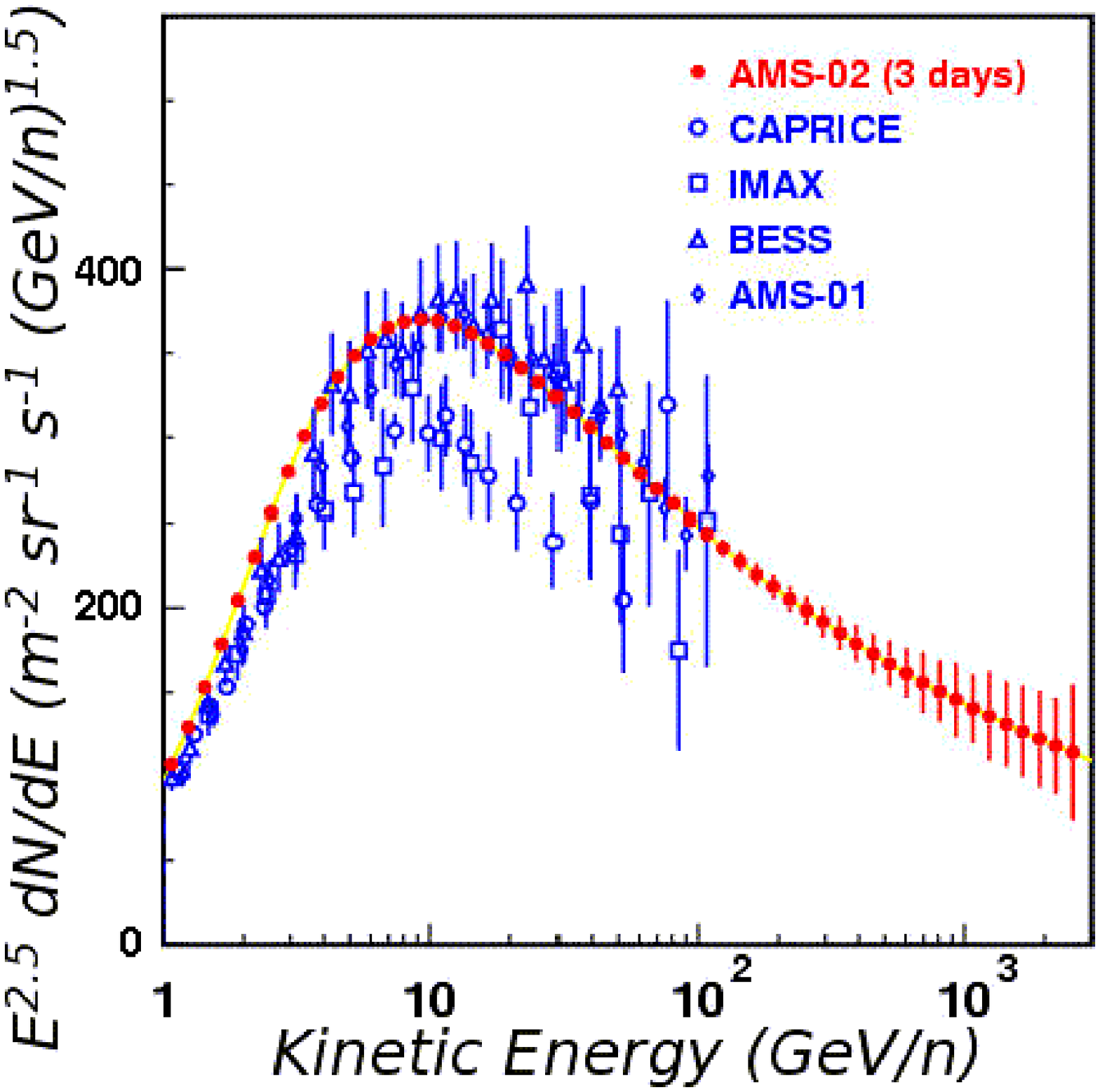}
\caption{CR proton (left) and helium (right) flux measurements are
  compared to AMS-02 expected statistics.}\label{fig6}
\end{figure}

\subsection{Local bubble and solar system neighborhood}

 The solar system is inside the ``local bubble'', a low-density region
 with scale length of 100 pc with irregular shape, that is the result
 of the explosion of several supernovae in the neighborhood \cite{maiz01}.
 Recent simulations \cite{maurin02} show that the ratio between
 unstable and stable isotopes is very sensitive to the features of the
 local bubble.  Figure~\ref{fig5}, left panel, shows the expected AMS-02
 statistics for the $^{10}$Be/$^9$Be measurement.

 One can probe larger distances measuring the electron and positron
 spectra, and the ratio between stable secondary species to primary
 nuclei, like the ratios of Li, Be and B to C.  In particular,
 electrons (figure~\ref{fig5}, right panel) can diffuse for relatively
 short time compared to CR nuclei, because of their large radiative
 losses.  High energy electrons (with $E \sim 1$ TeV) can reach
 distances of the order of 1 kpc, comparable with the galactic disk
 thicness, and lower energy electrons can propagate further beyond
 this limit: with electrons, the spherical propagation model becomes a
 rough approximation and a two-phases (disk and halo) cylindrical
 model is normally used.

\subsection{Galactic Interstellar Medium}

 A two-phases cylindrical model of the Galaxy is also used to simulate
 the propagation of protons and helium nuclei (figure~\ref{fig6}) in
 the interstellar medium, where they diffuse for roughly $2 \times
 10^7$ years.  Protons and helium nuclei are the deepest charged
 probes of the Galaxy: they diffuse on the average through one third
 of the Galactic disk and in the halo, before being measured
 \cite{maurin02}.  Other primary nuclei, like carbon, nitrogen, oxygen
 and iron, travel for long distances in the interstellar medium, where
 their interactions produce the secondary nuclei (Li, Be, B, and
 sub-Fe elements) that are not produced by the stellar
 nucleosynthesis, and whose fluxes are used to constrain the
 parameters of the galactic CR diffusion.  AMS-02 will provide very
 accurate measurements of all elements up to Fe, thus allowing for a
 deeper knowledge of the characteristics of the CR sources (supernova
 explosions) and the interstellar medium.

\end{document}